\documentclass[superscriptaddress,twocolumn,floatfix,preprintnumbers]{revtex4}
\usepackage[caption=false]{subfig} 
\usepackage{epsfig}
\usepackage{graphics}
\usepackage{graphicx}
\usepackage{amsmath,amssymb}
\usepackage{amsfonts}
\usepackage{bm}
\usepackage[usenames,dvipsnames]{color}
\usepackage{amssymb}
\usepackage{psfrag}
\usepackage{subfig}
\usepackage{url}
\usepackage{times}

\newcommand{\FF}{\mathrm{FF}}

\begin{document}

\title{Improving the sensitivity of a search for coalescing binary
  black holes with non-precessing spins in gravitational wave data}

\textcolor{red}{LIGO DCC: P1200132-v3}

\author{Stephen~Privitera} 
\email{sprivite@ligo.caltech.edu}
\affiliation{LIGO Laboratory, California Institute of Technology,
  1200 E. California Blvd., Pasadena, CA, USA}

\author{Satyanarayan~R.P.~Mohapatra}
\email{satya@astro.rit.edu}
\affiliation{Center for Computational Relativity and Gravitation and
  School of Mathematical Sciences, Rochester Institute of Technology,
  85 Lomb Memorial Drive, Rochester, NY 14623, USA}
\affiliation{Department of Physics, Syracuse University, Syracuse, NY
  USA}

\author{Parameswaran~Ajith}
\email{ajith@icts.res.in}
\affiliation{International Centre for Theoretical Sciences, 
Tata Institute of Fundamental Research, Bangalore 560012, India.}

\author{Kipp~Cannon}
\email{kcannon@cita.utoronto.ca}
\affiliation{Canadian Institute for Theoretical Astrophysics,
60 St. George St., Toronto, Canada.}

\author{Nickolas~Fotopoulos}
\affiliation{LIGO Laboratory, California Institute of Technology, 1200
  E. California Blvd., Pasadena, CA, USA}

\author{Melissa~A.~Frei}
\affiliation{Center for Computational Relativity and Gravitation and
  School of Mathematical Sciences, Rochester Institute of Technology,
  85 Lomb Memorial Drive, Rochester, NY 14623, USA}

\author{Chad~Hanna} 
\email{channa@perimeterinstitute.ca}
\affiliation{Perimeter Institute for Theoretical Physics,
31 Caroline Street North, Waterloo, Ontario, Canada}

\author{Alan~J.~Weinstein} 
\email{ajw@ligo.caltech.edu}
\affiliation{LIGO Laboratory, California Institute of Technology,
  1200 E. California Blvd., Pasadena, CA, USA}
  
\author{John~T.~Whelan}
\email{john.whelan@ligo.org}
\affiliation{Center for Computational Relativity and Gravitation and
  School of Mathematical Sciences, Rochester Institute of Technology,
  85 Lomb Memorial Drive, Rochester, NY 14623, USA}

\date{\today}


\begin{abstract}

\begin{center}


\end{center}

We demonstrate the implementation of a sensitive search pipeline for
gravitational waves from coalescing binary black holes whose
components have spins aligned with the orbital angular momentum. We
study the pipeline recovery of simulated gravitational wave signals
from aligned-spin binary black holes added to real detector noise,
comparing the pipeline performance with aligned-spin filter templates
to the same pipeline with non-spinning filter templates. Our results
exploit a three-parameter phenomenological waveform family that models
the full inspiral-merger-ringdown coalescence and treats the effect of
aligned spins with a single \textit{effective spin} parameter
$\chi$. We construct template banks from these waveforms by a
stochastic placement method and use these banks as filters in the
recently-developed {\tt gstlal} search pipeline.  We measure the
observable volume of the analysis pipeline for binary black hole
signals with $M_\mathrm{total} \in [15,25]M_\odot $ and $\chi \in
[0, 0.85]$. We find an increase in observable volume of up to 45\% for
systems with $0.2 \leq \chi \leq 0.85$ with almost no loss of
sensitivity to signals with $0\leq \chi\leq 0.2$. We demonstrate
this analysis on 25.9 days of data obtained from the Hanford and
Livingston detectors in LIGO's fifth observation run.

\end{abstract}

\maketitle


\section{Overview}
The coalescence (late-stage inspiral, merger, and ringdown) of binary
black hole (BBH) systems is one of the many promising sources of
gravitational waves expected to be detectable by ground based
detectors such as LIGO~\cite{Abbott:2007kv} and
Virgo~\cite{Accadia:2012zz}. The observation of such systems in
gravitational waves will serve as a direct probe of general relativity
in the strong-field, highly dynamical regime~\cite{testing_gr} and
have significant implications for the mechanism of their formation
through measurements of their coalescence rate as a function of their
masses and spins~\cite{lrr-2009-2}.

The components of a coalescing binary black hole system are expected
to have significant spins~\cite{gammie:black_hole_spin_theory,
reis:black_hole_spin_measurements,mcclintock:black_hole_spin_measurements}. Binary
black holes formed through ordinary binary stellar evolution from two
massive progenitor stars are expected to have spins that are nearly
aligned with their orbital angular momentum~\cite{lrr-2006-6}. On the
other hand, black hole binaries formed directly through dynamical
capture in dense stellar environments, such as globular clusters, will
typically have spins arbitrarily oriented relative to each other and
to the orbital angular momentum~\cite{Kalogera}. Which of these
processes contribute to the formation of binary black holes and by how
much is highly uncertain~\cite{ratesdoc}. Measuring the rate, masses
and spins of coalescing BBHs therefore will directly inform the
processes by which these system form, the rate at which these
processes proceed and in the case of common binary evolution, the
properties of progenitor systems which eventually become BBHs.

The component spins of a binary black hole system are encoded in the
gravitational wave signal emitted during
coalescence~\cite{PhysRevD.49.6274} and are measurable from the
observation of these signals~\cite{s6_parameter_estimation}. Binary
systems with arbitrarily oriented component spins undergo precession
of the orbital plane. This precessional behavior manifests in the
gravitational wave signal as a quasi-periodic modulation of the signal's
amplitude and phase. On the other hand, binaries with spins aligned
with the orbital angular momentum do not exhibit precession and no
such modulation is present in their gravitational wave
signal. Instead, aligned spin systems exhibit non-periodic (or
``secular'') contributions to the amplitude and phase of their
gravitational wave signals and these effects make possible, in
principle, the discrimination between aligned spin and non-spinning
systems. Since the magnitude and orientation of the spins relative to
the orbital angular momentum is determined by the processes that
create these binary systems, distinguishing between these three cases
(precessing, non-precessing and non-spinning) in gravitational wave
observations allows for the interpretation of these observations in
terms of the astrophysical mechanism which creates these systems.

With the availability of predictions for the waveforms emitted during
coalescence, searches for gravitational waves from BBHs profit greatly
from the use of matched filtering to sieve the data. The matched
filter gives the maximal signal-to-noise ratio (SNR) among all
possible linear filters in the ideal case of a Gaussian-distributed
background, assuming that the template corresponds exactly to the
signal potentially in the data. In practice, we do not know {\it a
priori} the exact parameters of the binary system which creates the
signal and we therefore analyze the data using a large number of
waveforms with discretely sampled parameters. The true signal
typically will not be identically one of the filter templates, either
because of the discreteness of the bank or because of imperfect
waveform modeling, in which case the matched filtering technique
becomes sub-optimal. However, if the true signal is similar enough to
one of the filter templates, the expected loss of SNR due to these
effects may be negligibly small.

The data analysis problem is significantly complicated by non-Gaussian
artifacts present in gravitational wave data from real detectors. In
this case, the matched filter SNR is not optimal for detection and we
require additional background-signal discrimination techniques. For
example, recent LIGO and Virgo
searches~\cite{Collaboration:S5HighMass, 2013PhRvD..87b2002A,
Collaboration:S6CBClowmass} have benefited greatly from the use of a
$\chi^2$ statistic~\cite{Allenfg:2004}, which uses the triggered
template to subtract out the putative signal from the data and test
whether the resulting data stream is consistent with Gaussian
noise. As with the matched filter, the efficacy of this statistical
test relies crucially on having an accurate waveform model for the
signal. The $\chi^2$ test can easily mistake a large mismatch between
the template and the signal as being due to non-Gaussian noise and the
event could consequently be missed by the search.

Given that BBH systems probably have significant spins and that the
search sensitivity depends strongly on having accurate waveform
models, the inclusion of spin effects in search templates has for good
reason been a long-standing goal in the
field~\cite{BuonannoChenVallisneri:2003b, S3_BCVSpin,
VanDenBroeck:2009gd, Pan:2003qt, HarryFairhurst:single_spin,
Fazi_thesis:2009, brown_lund_1, brown_lund_2, brown_lund_3,
2012arXiv1210.6666A}. Yet none of the previous attempts to include
spin effects in templates have resulted in improved search
sensitivity. In~\cite{S3_BCVSpin}, an analysis of data from LIGO's
third science run demonstrated for the first time the use of spinning
templates~\cite{BuonannoChenVallisneri:2003b} in an actual search.
However, it was later shown~\cite{VanDenBroeck:2009gd} that the
pipeline used for this search with spinning templates was no more
sensitive to spinning systems than the same analysis using
non-spinning templates, even though the spinning templates recovered
more signal-to-noise for simulated spinning signals.  The results of
these studies motivated the decision to neglect spin effects in search
templates in the most recent LIGO and Virgo searches for BBH
coalescence~\cite{LIGOS3S4all,Collaboration:S5HighMass,
2013PhRvD..87b2002A, Collaboration:S6CBClowmass} and to date no other
search of LIGO and Virgo data has used spinning templates.

The null result in~\cite{S3_BCVSpin} was attributed to the elevation
in the rate of background events owing to the large number of
parameters required to describe the spinning template waveforms. It
was immediately realized that to make the use of spinning templates in
an analysis beneficial, better signal consistency tests would have to
be developed and implemented to suppress the increased background
event rates. In particular, the highly effective $\chi^2$ test used in
the contemporary LIGO search with non-spinning
templates~\cite{LIGOS3S4all} was never integrated into the spinning
search pipeline and probably at a severe cost for the results of the
sensitivity analysis. Following this work, two other pipelines were
developed~\cite{HarryFairhurst:single_spin,Fazi_thesis:2009} which
included spinning templates based on a modified phenomenological model
for single-spin binaries~\cite{Pan:2003qt}, but neither analysis
demonstrated conclusively that the techniques would improve the search
sensitivity. Neither of these investigations examined the use signal
consistency tests for suppressing the background.

A couple of pipeline implementation issues arose from these
investigations, which ultimately concern the question of having a
measure for the ``distance'' between nearby templates. The first was
the problem of defining coincidence between triggers in different
detectors. In the two-dimensional mass parameter space, elegant and
rigorous techniques exist for defining when the parameters of two
triggers are close enough to be considered the
same~\cite{ref:ethinca}. However, it was not clear at the time how to
extend this method to define a robust coincidence criterion in
higher-dimensional parameter spaces. The analyses in~\cite{S3_BCVSpin}
and~\cite{Fazi_thesis:2009} defined coincidence between triggers in
terms of the standard mass coincidence criterion proposed
in~\cite{ref:ethinca} together with a simple interval cut on the
remaining parameters. The study in~\cite{HarryFairhurst:single_spin}
was for a coherent analysis for which the question of coincidence is
irrrelevant.

Another problem that comes with having a larger template parameter
space is that of efficiently placing templates to minimize the loss of
SNR arising from the discreteness of the template bank. Recent LIGO
and Virgo compact binary searches, which used non-spinning templates,
have relied on a lattice placement technique known to select the
fewest number of templates for a given tolerance of SNR
loss.~\cite{hexabank,Owen:1995tm,Owen:1998dk,Sathyaprakash:1994nj}.
This technique relies on having knowledge of certain special
parameters for which the mismatch between two neighboring templates is
the ordinary Euclidean norm applied to these parameters. It is then
straightforward to construct a regular lattice in these coordinates to
guarantee a specified maximal loss of SNR for any signal which lies
between the templates. However, the method requires one to determine
which coordinates, if any, are appropriate for the lattice
construction. In larger dimensional parameter spaces or when using
waveforms which include effects beyond the inspiral portion of the
coalescence, we often do not have any known special parameters in
which to form the lattice and other approaches to template placement
become necessary.

In the studies discussed above, two main approaches were taken to
construct banks of spinning templates. The first is a simple stacking
method in which one lays out a grid of points in the spin parameters
and for each point in the grid applies to the lattice technique to lay
out templates in $m_1$ and $m_2$. The main limitation of this
technique is that it is not known {\it a priori} how fine the grid
spacing needs to be to fully cover the space. One determines the
required spacing through simple trial and error. Additionally, this
approach provides very little assurance that the resulting template
bank is close to the minimal size bank needed to cover the space.

In~\cite{S3_BCVSpin}, the authors also explored the use of a
stochastically generated bank and found it to give significantly fewer
templates than the stacking approach. More recently, several groups
have conducted thorough and systematic studies of the stochastic
template placing techniques based~\cite{stochastic1,cover-art} and the
outlook is quite promising~\cite{2012arXiv1210.6666A}. Stochastic placement techniques are
applicable to a wide variety of waveform approximants, requiring no
prior knowledge of special parameters and extending straight-forwardly
to higher dimensions. We discuss these template placement methods in
more detail in the following section.

Here we revisit the problem of using spinning waveform models as
search templates. Our analysis uses the three-parameter IMRPhenomB
waveform family~\cite{IMRSAPaper} modeling the inspiral, merger and
ringdown of binary black hole systems with aligned spins. This
waveform family captures the dominant effects of aligned spin with a
single \textit{effective spin} parameter $\chi$ defined below.  We
construct template banks from these waveforms using a generic,
extensible stochastic placement infrastructure ${\tt
SBank}$~\cite{2012arXiv1210.6666A} implemented in the {\tt LAL}
gravitational wave data analysis library~\cite{LAL} and incorporate the ${\tt
SBank}$ infrastructure into the recently-developed ${\tt
gstlal}$ pipeline~\cite{gstlal_lloid}. We then apply the search
pipeline to 25.9 days of detector noise obtained from the initial LIGO
detectors at Hanford and Livingston. We demonstrate for this data set
an increase in the sensitive volume for signals with $\chi\in [0.2,0.85]$ of up to 45\%
with only a moderate loss of sensitivity to signals with $\chi\in [0,0.2]$ when
the filter templates include the effects of aligned spin.

In a few years, the advanced LIGO and Virgo detectors will come online
with much better sensitivity compared to initial LIGO
detectors~\cite{0264-9381-27-8-084006,advanced_virgo}.  As these
detectors will have significantly improved sensitivity and bandwidth,
the importance of including spin in waveform templates will only
become greater. While advanced LIGO and Virgo may bring with them
non-Gaussian backgrounds unlike those seen in initial LIGO, we
emphasize that our results here for the improved sensitivity to
spinning signals are demonstrated in real detector data from initial
LIGO and in doing so directly confronts the hazards of realistic
non-Gaussian detector behavior. The template placement infrastructure
and analysis pipeline we use in this work are readily scalable to the
data analysis requirements of advanced LIGO and Virgo sensitivities
and we anticipate that the improvements shown here will be readily
transferable to BBH searches in advanced LIGO and Virgo data.

In the following section, we briefly review commonly used template
pacement strategies and motivate our adoption here of the stochastic
approach. We then apply the stochastic placement method to construct
aligned-spin and non-spinning template banks. In doing so, we also
identify the regions of parameter space with the greatest potential
for improvement in sensitivity by the inclusion of spin. Focusing in
on these regions, we then demonstrate the implementation of our
template banks in the {\tt gstlal} search pipeline. We find for a
fixed false alarm rate that the pipeline analysis with aligned-spin
templates has an observable volume that is 95\% to 145\% that of the
observable volume for same pipeline with non-spinning templates in the
considered parameter space. Furthermore, we show that the use of
aligned-spin templates allows for more accurate mass and spin
parameter recovery in the pipeline. We conclude this paper with a
discussion of other compact binary search problems to which we hope
the methods detailed here will apply.

\section{Stochastic Template Placement}
\label{sec:sbank}
As mentioned in the introduction, a matched filter search for binary
black hole coalescences requires the use of a bank of template
waveforms with discretely sampled source parameters. In this section,
we review lattice and stochastic template placement techniques and
define the notions we will use in the subsequent section to
quantitatively describe the efficiacy of a given template bank for the
detection of a population of target signals

A useful quantitative measure for the effectiveness of a template bank
towards capturing the SNR for a target signal population is
the \textit{fitting factor}~\cite{Apostolatos:1995}, which we now
define. Given two waveforms $h_1(t)$ and $h_2(t)$, we define the
overlap between two waveforms $h_1(t)$ and $h_2(t)$ by the integral
\begin{equation}
\left< h_1 | h_2 \right> = 2 \int_{f_\mathrm{low}}^\infty \frac{\tilde{h}_1(f)\tilde{h}^*_2(f) + \tilde{h}^*_1(f)\tilde{h}_2(f)}{S_n(f)} df,
\end{equation}
where $S_n(f)$ is the (one-sided) power spectral density of the
detector noise, $f_\mathrm{low}$ is the low frequency cutoff, the
tilde denotes the Fourier transform of the waveform and the asterisk
denotes complex conjugation~\cite{Allen:2005fk}. We denote normalized
waveforms with a hat so that
$\langle \hat{h}_i|\hat{h}_i\rangle=1$. If we denote the template bank
by $B= \{\hat{h}_i\}_{i=1}^N$ and $\hat{h}_{\vec{\lambda}}$ is some
target signal with arbitrary source parameters $\vec{\lambda}$, then
we define \textit{fitting factor} of the bank towards this signal by
\begin{equation}
\FF(\vec{\lambda}; B) = \max_{i, t} \left<\hat{h}_i
| \hat{h}_{\vec{\lambda}} \right>,
\end{equation}
where the maximization is taken not only over the templates in the
bank but also over the time translation of the templates.  In the
construction of a template bank, the goal generally is to achieve the
highest possible fitting factors with the fewest possible templates.

Currently implemented LIGO-Virgo matched filtering searches for BBHs
are based on a lattice approach for placing the
templates~\cite{Collaboration:2009tt,cbc-highmass} and similar
techniques are being developed for advanced generation
searches~\cite{brown_lund_1, brown_lund_2, brown_lund_3}. This
technique relies on the existence of coordinates
$\vec{\lambda}'=f(\vec{\lambda})$ such that at any point
$\vec{\lambda}$ in the source parameter space the fitting factor
between the bank $B$ and the signal $h_{\vec{\lambda}}$ is
approximately
\begin{equation}
\FF(\vec{\lambda};B) \approx 1 - \min_i |f(\vec{\lambda}_i) - f(\vec{\lambda})|^2.
\label{eqn:metric_match}
\end{equation}
for $|\vec{\lambda}_i - \vec{\lambda}| << 1$. One can then place the
templates on a regular lattice in these coordinates to guarantee a
minimal loss of SNR~\cite{Owen:1998dk,hexabank}. The lattice technique
is highly computationally efficient but requires one to determine the
appropriate coordinates in which to lay down the templates.

In the inspiral-merger-ringdown regime, however, no such coordinates
are known and the lattice techniques cannot be reliably applied to
place templates. Therefore, in this study we adopt a stochastic
approach~\cite{stochastic1,cover-art} to place templates in the mass
and spin parameter space of binary black hole systems. In this
approach, template parameters are proposed randomly and the proposed
waveform's fitting factor is computed against some initial seed
bank. A threshold value for the minimal fitting factor is chosen (here
we take $FF_\mathrm{min} = 0.97$) and if the proposed template does
not achieve a fitting factor larger than this value, then the proposal
is added to the bank. This extended bank becomes the seed for the next
iteration. Otherwise, the proposal is discarded and the same bank is
used for the seed of next iteration. The iteration continues until the
rejection rate of proposals becomes sufficiently high.

We reuse a recently-developed generic and extensible impementation of
this algorithm, referred to as {\tt SBank}, which has previously
proven effective in constructing template banks for lower mass compact
binary systems with spin~\cite{2012arXiv1210.6666A}. The stochastic
placement technique is robust and we show here that it works well in
three dimensions. While the implementation of this technique to higher
dimensions in straightforward, the required convergence time for
generating a higher-dimensional bank may become impractical and the
resulting template banks can be significantly larger than the
theoretically optimal template bank
size~\cite{cover-art}. Furthermore, even in a small number of
dimensions, long waveforms can lead to computationally costly FFTs and
considerably slow down the convergence time. In such cases, it is
possible to side-step the numerical fitting factor calculation with an
analytic approximation to the fitting factor as in
Eqn.~\ref{eqn:metric_match} and demonstrated in
~\cite{2012arXiv1210.6666A}.

\section{Template Bank Effectualness}
\label{sec:effectualness}
Here we determine the regions in parameter space which have the
greatest potential for improvement in SNR recovery by the inclusion of
aligned-spin waveforms in a search template bank. Our study employs
the IMRPhenomB waveform family~\cite{IMRSAPaper}, which models the
late-inspiral [that is, $f_\mathrm{GW} \gtrsim
10^{-3}/(GM_\mathrm{total}/c^3)$, where $f_\mathrm{GW}$ is the
dominant mode gravitational wave frequency], merger and ringdown
stages of the coalescence of binary black holes with aligned
spins. The waveform family is parametrized by the component masses
$m_1$ and $m_2$ of the binary and a single \textit{effective spin}
parameter $\chi$. The effective spin parameter captures the
dominant waveforms effects arising from aligned spin and reduces the
dimensionality of the intrinsic parameter space from four to three. The effective spin is defined by
\begin{equation}
    \label{eqn:chidef}
    \chi \equiv \frac{m_1\chi_1+m_2\chi_2}{m_1+m_2},
\end{equation}
where $\chi_1$ and $\chi_2$ are the dimensionless spins of the
component black holes.

The IMRPhenomB waveform model consists of a parametrized
phenomenological fit to hybrid waveforms constructed from numerical
relativity simulations of the late-inspiral, merger and ringdown of
binary black holes matched to a post-Newtonian approximation
describing the early inspiral. As such, the validity of these
waveforms have restrictions on the mass ratio and spins based on the
availability of numerical simulations with which to fit. Specifically,
the IMRPhenomB family is expected to be accurate only for low to
moderate mass ratios and spins.  Hence, in this study, we consider
only binaries for which $1 \le m_1/m_2 \le 4$ and $-0.5 \le \chi <
0.85$.

\begin{figure}

        \centering
        \subfloat[$\chi = 0$ Template Bank \label{fig:nonspin_effectualness}]{%
                \includegraphics[width=0.5\textwidth]{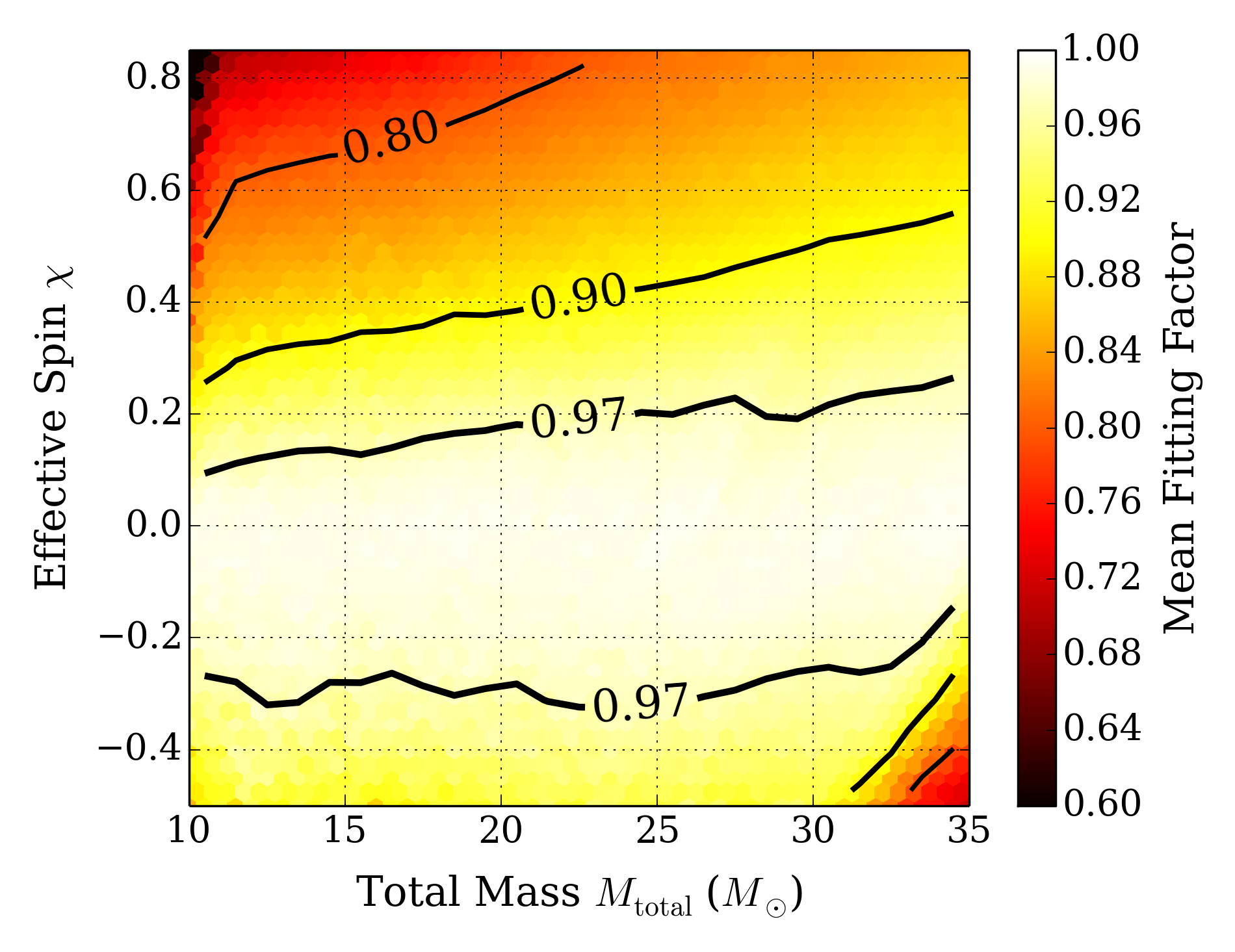}}

        \centering
        \subfloat[$\chi \ge 0$ Template Bank \label{fig:positive_spin_effectualness}]{%
                \includegraphics[width=0.5\textwidth]{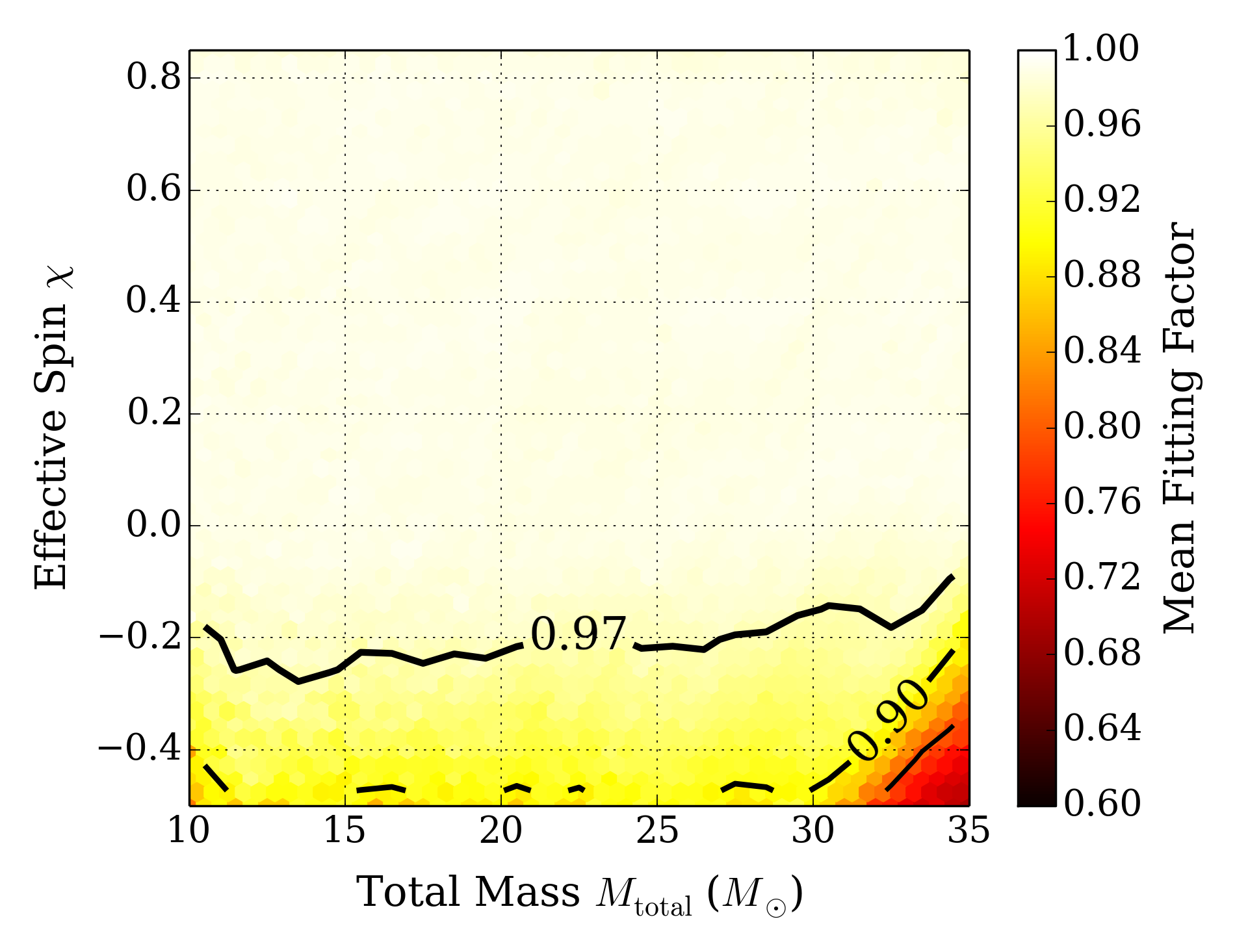}}

\caption{\textit{Capturing aligned-spin effects in template banks.} %
  Above we show the expected fractional signal-to-noise recovery for a
  population of aligned-spin binary black holes using a bank of
  IMRPhenomB waveforms with (a) $\chi=0$ and (b) $\chi \ge
  0$. The solid lines indicate the approximate fitting factor contours
  in the $M_\mathrm{total}-\chi$ plane, averaging over the mass
  ratio dimension with $1 \le m_1/m_2 \le 4$. The template banks are
  both constructed with the stochastic placement method described in
  Section~\ref{sec:sbank} assuming the design iLIGO
  sensitivity~\cite{LIGO-E950018-02-E} with $f_\mathrm{low}=40$Hz.  We
  find that with this sensitivity, a template bank that neglects spin
  achieves fitting factors exceeding the nominal $FF_\mathrm{min} =
  0.97$ from aligned-spin systems over a wide region of parameter
  space, spanning roughly $-0.25 \le \chi \le 0.2$ over the entire
  mass range. As the mass of the system increases, the loss of
  signal-to-noise incurred from neglecting spin becomes small and we
  therefore do not consider systems with total masses exceeding
  $M_\mathrm{total} = 35~M_\odot$. The $\chi=0$ bank has $\sim
  700$ templates, whereas the $\chi \ge 0$ bank has $\sim 3000$
  templates.}

\label{fig:effectualness}
\end{figure}

We choose to further focus only on the regions in the parameter space
where the merger and ringdown stages are important for detection.  For
an initial LIGO design sensitivity, the effects of merger and ringdown
begin to contribute significantly to the SNR when the total mass of
the binary exceeds $M_\mathrm{total}\approx
12~M_\odot$~\cite{BuonannoIyerOchsnerYiSathya2009}. For lower mass
systems, accurate and generically spinning post-Newtonian waveforms
are available~\cite{LAL} and can be used to give a more
detailed understanding the effects of spin on the search. We therefore
consider only systems with $M_\mathrm{total}\geq 10~M_\odot$, giving a
small safety factor between the transitional region and considering
the degeneracy between the mass and spin parameters.

Since the finite size of neutron stars can have a significant impact
on the gravitational waveform observed in the merger phase of
coalescence, we restrict our attention to binary black holes and take
$m_i = 3~M_\odot$ as the minimal component mass. We note that from
astrophysical considerations, neutron stars in coalescing compact
binaries are not expected to have large spins. Further, from physical
considerations of the possible neutron star equations of state, the
dimensionless spin for a neutron cannot exceed $\sim0.7$ without
undergoing tidal distruption.

\begin{figure}
        \centering \includegraphics[width=0.45\textwidth]{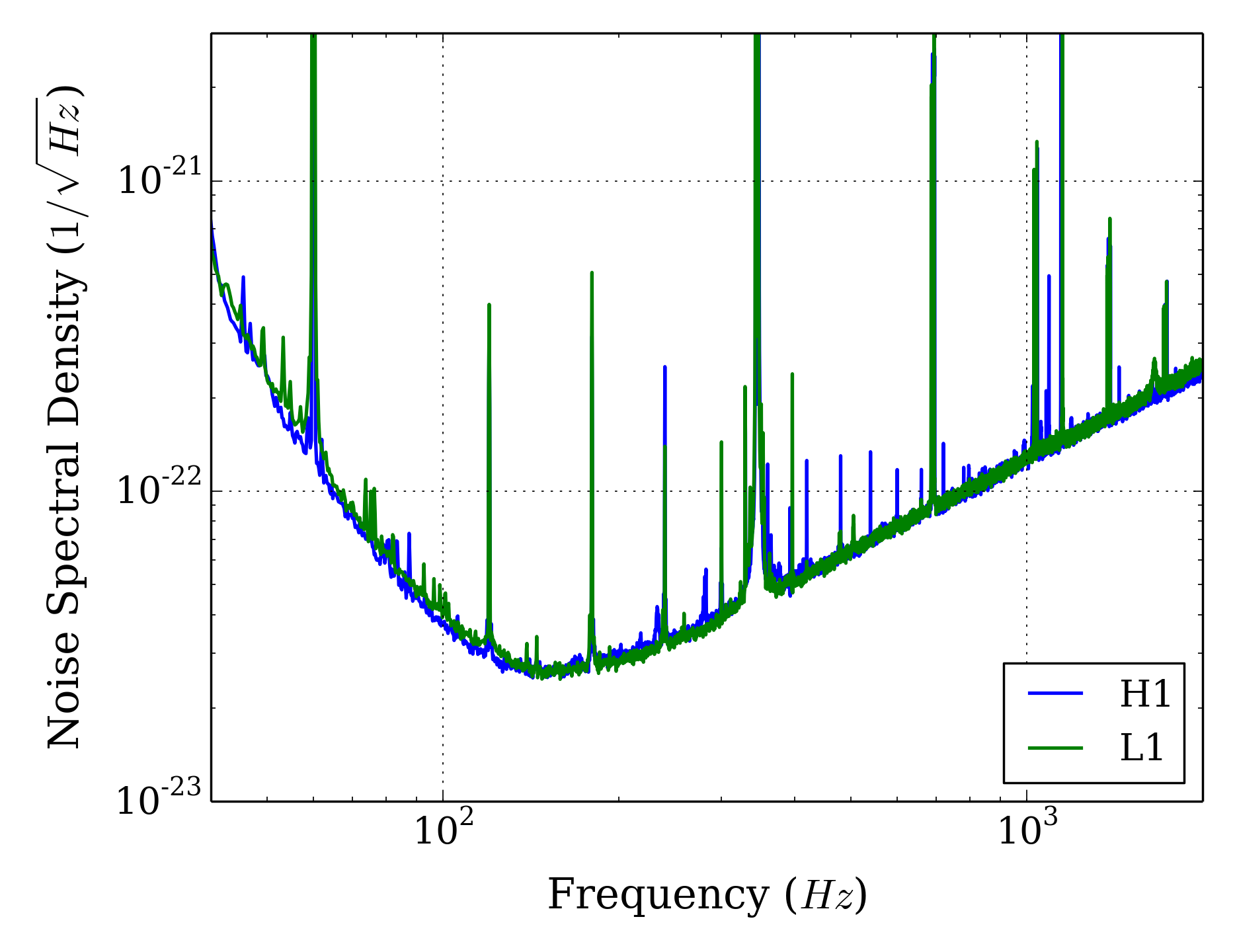}

        \caption{\textit{Detector sensitivities.} Here we show
          characteristic noise spectral density curves for data from
          the S5 observational run of the two detectors H1 and L1 used
          to supply noise to which simulated signals are added
          in this study~\cite{sensitivity_doc_s5}.}
\label{fig:psd}
\end{figure}

We used the {\tt SBank} implementation of the stochastic placement
method described in Section~\ref{sec:sbank} to generate a bank of
IMRPhenomB templates with $\chi=0$ using the above mass parameter
restrictions and $f_\mathrm{low}=40$Hz. We then computed the fittings
factors of this template bank towards aligned spin signals in the same
$m_1-m_2$ parameter space.  In Fig.~\ref{fig:nonspin_effectualness},
we show that a template bank with $\chi=0$ already captures greater
than 97\% of the possible SNR over a wide mass and spin range. In
particular, we note that the $\chi=0$ bank covers signals with $\chi <
0$ down to roughly $\chi \sim -0.25$ over the entire mass range. From
astrophysical considerations of binary evolution, spins positively
aligned with the orbital angular momentum are considered the more
likely scenario for binary black holes~\cite{lrr-2006-6}. Given these
factors, along with the potential for artifacts in the waveforms at
large negative $\chi$, we develop our search using only $\chi \ge 0$
templates. Note that since $\chi$ is a mass-weighted sum of the two
component spins, this restriction does not necessarily exclude the
possibility that one of the black holes has an anti-aligned spin. We
also see that as the total mass of the target system increases, the
fractional loss of SNR incurred from neglecting spin decreases. This
effect is due to the fact that higher mass systems merge at lower
frequencies and have fewer cycles in the LIGO sensitive band and
consequently the matched filtering is more tolerant of imperfect
templates. We thus expect that for systems with total masses exceeding
$M_\mathrm{total} = 35M_\odot$, the benefits of including spin effects
will be small.

In Fig.~\ref{fig:positive_spin_effectualness}, we demonstrate the
coverage of the parameter space obtained by including only waveforms
for non-negative aligned spins ($\chi \ge 0$) in the template
bank. The improvement in SNR recovery obtained by using such a bank
comes at the cost of having more than three times as many templates in
the bank. For the non-spinning case, we constructed a bank with $\sim
700$ templates, while to cover the positively aligned signals, we
require $\sim 3000$ templates. The increase in the number of templates
will increase the number of background triggers and detecting a signal
at a given false probability requires raising the SNR threshold used
for detection.

The expected SNR scales inversely with distance to the source and the
detection rate scales as the cube of the sensitive distance.
Therefore, the fractional increase in detection rate due to the
inclusion of spinning waveforms in the template bank should be given
by
\begin{equation}
\frac{V_\mathrm{spin}}{V_\mathrm{non spin}} = \left( \frac{\FF_\mathrm{spin}}{\FF_\mathrm{non spin}} \right)^3,
\end{equation}
where $\FF_\mathrm{non spin}$ and $\FF_\mathrm{spin}$ are the fitting
factors for the non-spinning and spinning template banks respectively
to the target signal. These statements regarding the increase in
detection rate assume that the only gain in sensitivity comes from the
SNR. In practice, our data also contain non-Gaussian artifacts and SNR
is not an optimal detection statistic, as mentioned above.

The characteristics of the background can change in complicated ways
when new template waveforms are introduced to a search. The results
presented in Fig.~\ref{fig:effectualness} do not reflect the impact of
non-Gaussianity in the data, nor to they capture the effects of
multi-detector coincidence requirements, the use of $\chi^2$
statistics, increased false alarms due to larger template banks or
other effects which are important in realistic search pipelines.In the
following section, we describe the implementation of these spinning
template banks in a search pipeline. We show that even in non-Gaussian
data, we are able to sufficiently suppress the extra background to
achieve a net gain in the search sensitivity.

\section{Implementation of Spin Effects in a Search of Gravitational Wave Data}
We now demonstrate the use of the template banks constructed in
Section~\ref{sec:effectualness} as filters in the {\tt gstlal} search
pipeline~\cite{gstlal_lloid}. Using these template banks, we measured
the sensitivity of the pipeline to a simulated population of more than
200000 binary black holes. Here we compare the mean sensitive
distance of the pipeline analysis when using a bank of aligned-spin
templates against that of an otherwise identical analysis which uses
non-spinning templates. Our simulated binary black hole systems were
populated with a uniform distribution in mass ratio, total mass and
effective spin with $m_1/m_2 \in [1,4]$, $M_\mathrm{total}\in
[15,25]M_\odot$ and $\chi \in [0, 0.85]$. As with the templates, the
simulated waveforms were computed using the IMRPhenomB
approximation. We conducted our study on 25.9 days of coincident
detector noise obtained from observations of the Hanford and
Livingston detectors during LIGO's fifth science run. Typical strain
sensitivities for these two detectors during this science run are
shown in Fig.~\ref{fig:psd}.

\begin{figure*}

        \centering
        \subfloat[Injections with $0 \leq \chi \leq 0.2$ \label{fig:relative_volumes_low}]{%
                \includegraphics[width=0.45\textwidth]{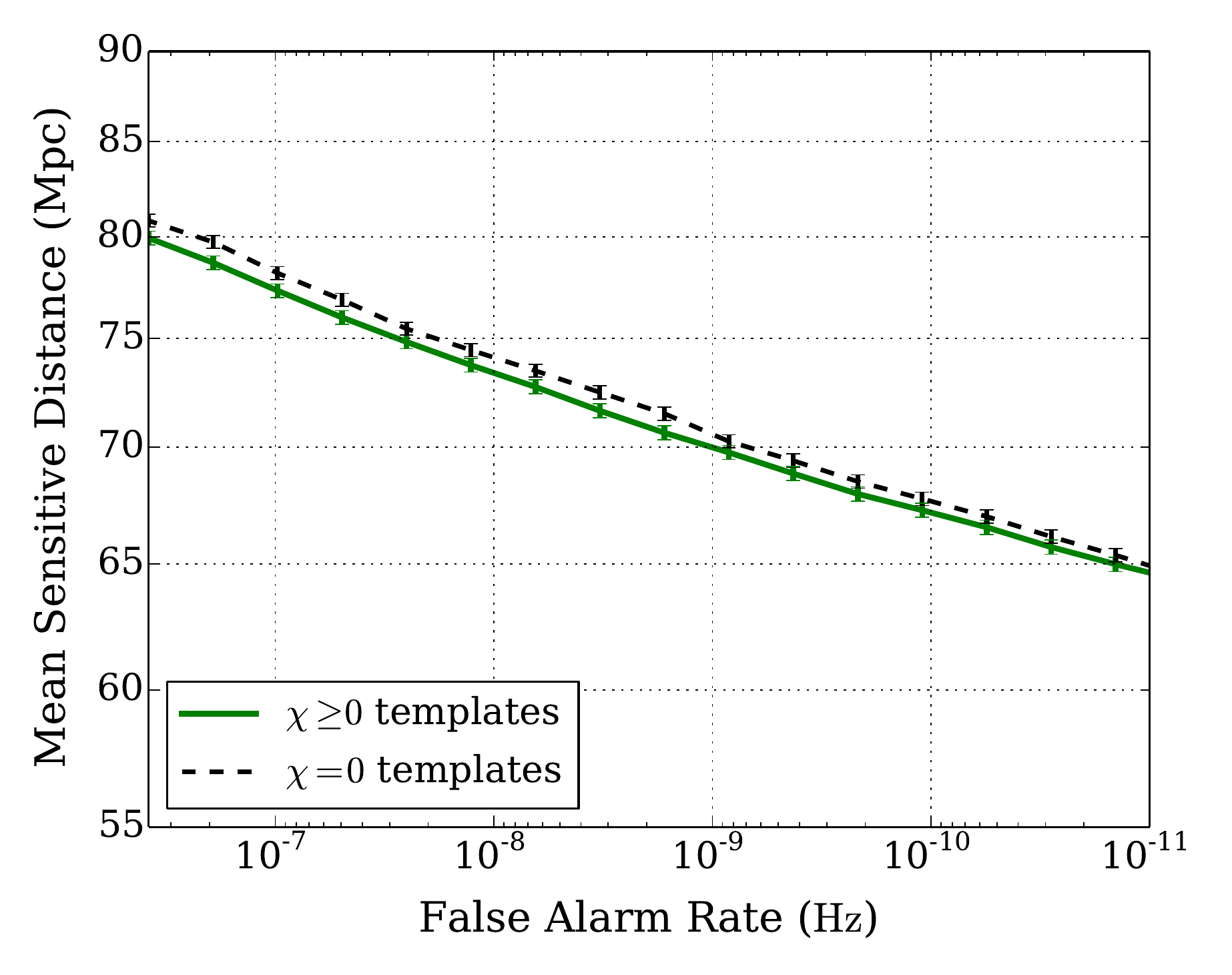}}
        \subfloat[Injections with $0.2 \leq \chi \leq 0.5$ \label{fig:relative_volumes_mid}]{%
                \includegraphics[width=0.45\textwidth]{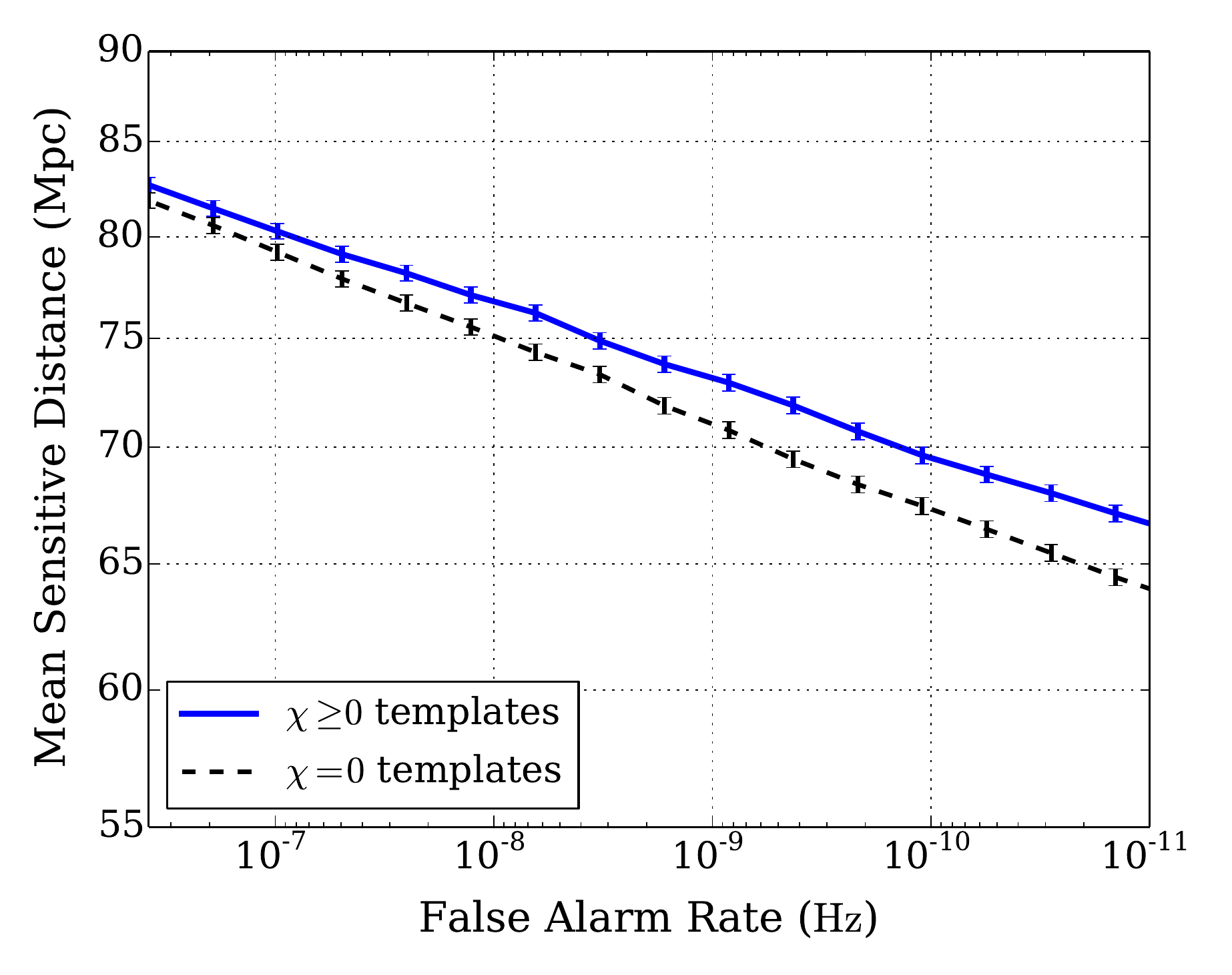}}\\

        \subfloat[Injections with $0.5 \leq \chi \leq 0.85$ \label{fig:relative_volumes_high}]{%
                \includegraphics[width=0.45\textwidth]{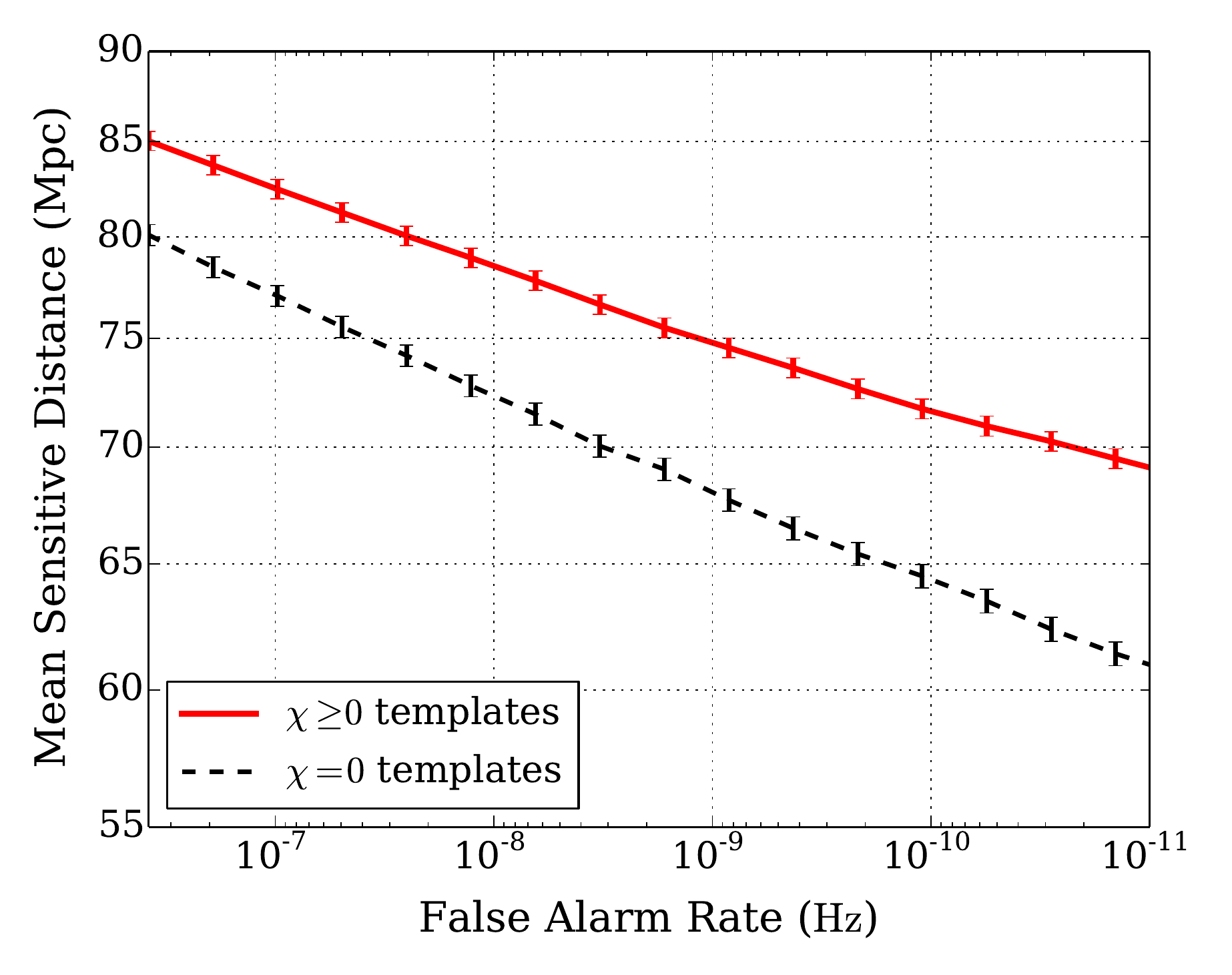}}
        \subfloat[Volume Ratios \label{fig:relative_volumes}]{%
                \includegraphics[width=0.45\textwidth]{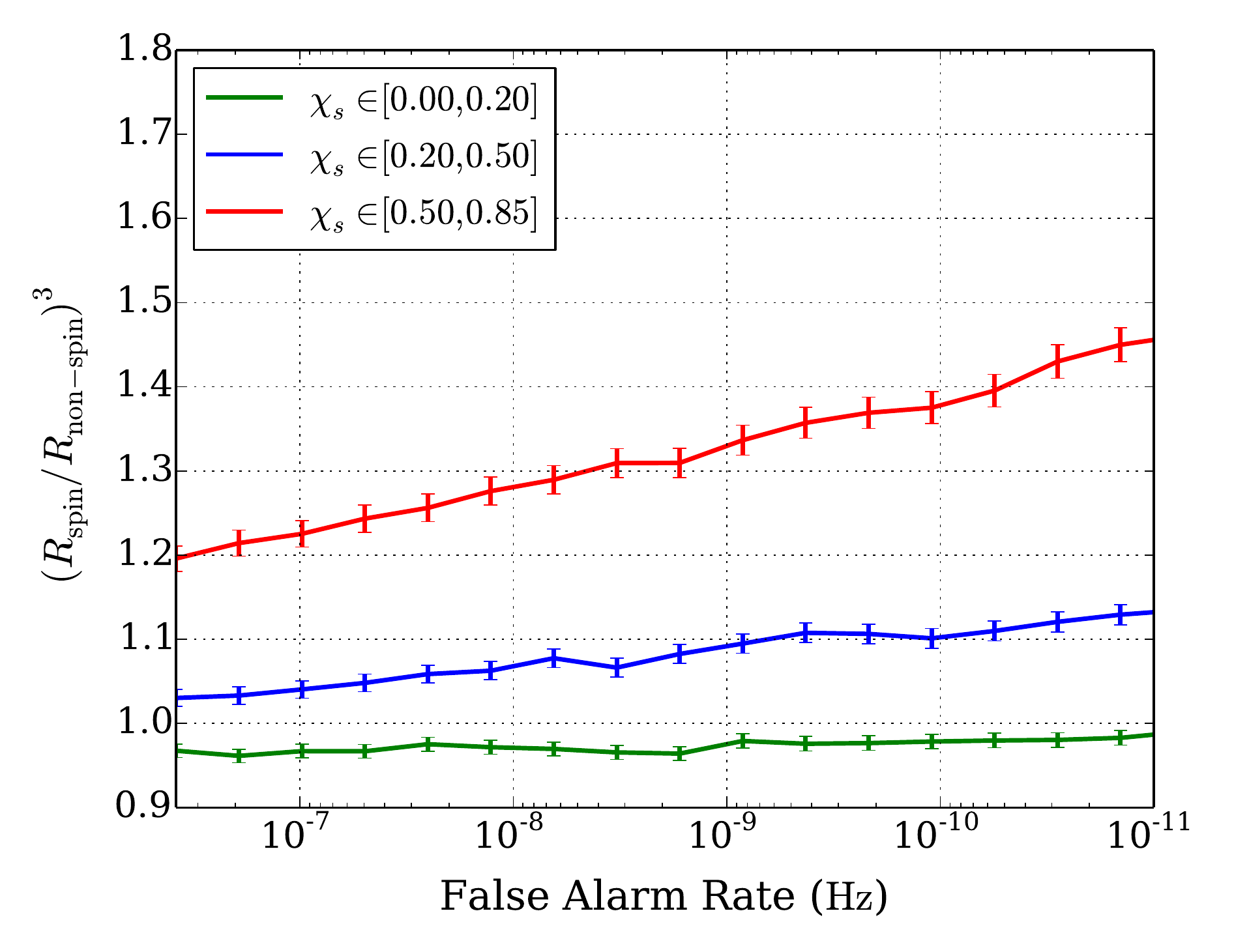}}

        \caption{\textit{Comparison of search sensitivities as a
            function of false alarm rate threshold.} Here we compare
            the sensitivities to aligned spin systems with
            $M_\mathrm{total} \in [15,25]M_\odot$ for an analysis
            which used templates with $\chi \geq 0$ and an analysis
            which used templates with $\chi=0$. The template banks
            each covered the mass range $M_\mathrm{total} \in
            [10,35]M_\odot$. In (a-c), we show the absolute
            sensitivities for these analyses in terms of the average
            distance to which the analyses identify an injection with
            a trigger above a given false alarm rate threshold. In
            (d), we show the ratios of the sensitive volumes for each
            of the three spin bins. We find that for injections with
            $\chi \geq 0.2$, the spinning search observes a larger
            sensitive volume than the non-spinning search for all
            false alarm rates by as much as 45\%. For injections with
            $0 \leq \chi \leq 0.2$, we observe a small but
            statistically significant decrease in sensitive volume on
            the order of 5\% incurred by the use of spinning
            templates.}

\label{fig:sensitivities}
\end{figure*}

In Fig.~\ref{fig:sensitivities}, we show the measured sensitivities of
our two analyses in terms of the mean distance accessible to each
search as a function of the false alarm rate threshold. We show our
results only for systems with total masses in the range
$M_\mathrm{total} \in [15,25]M_\odot$ to avoid complications
associated with the boundaries of the template banks, which covered
the range $M_\mathrm{total}\in [10,35]M_\odot$. As expected and
demonstrated in Fig.~\label{fig:relative_volumes_high}, we find that
the greatest improvement in sensitivity is for target systems with
high effective spins. As seen in Fig.~\ref{fig:relative_volumes}, the
volume improvement, and therefore the increase in detection rate, can
be as high as 45\% for these highly spinning systems. We emphasize the
non-trivial result shown in Fig.~\ref{fig:relative_volumes_low} that
for weakly-spinning target systems ($\chi\leq 0.2$), the analysis
with spinning templates and the analysis with non-spinning templates
have comparable sensitivities, with the aligned-spin template analysis
achieving at worst 95\% of the sensitive volume of the non-spin
template analysis. The apparent loss of detection rate in the small
effective spin regime is only applicable if we are wrong in our
expectations that black holes have significant spins.  Otherwise, we
expect this search method to increase the overall detection rate of
spinning BBH systems, provided that these spins are aligned. These
results demonstrate for the first time an analysis of real detector
data which is made more sensitive to spinning signals by the use of
spinning templates compared to the same analysis performed with
non-spinning templates.

The analysis performed here differs from previous attempts towards the
inclusion of spin effects in search templates in a number of
ways. Firstly, this analysis makes use of a template family that
captures the effect of non-precessing spins by using a small number
of \emph{physical} parameters, which allows us to construct a simple
three-dimensional template bank. Recent studies have suggested that
such template banks are effectual for a significant fraction or
precessing binaries as well~\cite{IMRSAPaper, 2012arXiv1210.6666A,
2011PhRvD..84h4037A}. This is in sharp contrast with the earlier work,
which either used phenomenological parameters to capture spin
effects~\cite{S3_BCVSpin, BuonannoChenVallisneri:2003b} or methods to
maximize the SNR over a number of extrinsic parameters that produced
elevated background~\cite{Pan:2003qt}.

This analysis also used an autocorrelation $\chi^2$ statistic,
analogous to the time-frequency statistic $\chi^2$ developed
in~\cite{Allenfg:2004} used in recent LIGO and Virgo
compact binary searches~\cite{Collaboration:S5HighMass,
2013PhRvD..87b2002A, Collaboration:S6CBClowmass}. The autocorrelation
statistic is based on the principle that the SNR time series obtained
from filtering data which contains a signal against a template that
closely matches the signal is approximately equal to the
autocorrelation function of the template plus noise. Subtracting the
template autocorrelation from the SNR time series and computing the
residual noise power gives a measure of the consistency of that data
with the signal model.

As discussed in the introduction, previous studies on the inclusion of
spin effects in template waveforms suffered in part due to the lack of
a sufficiently strong signal consistency tests to reject triggers
occurring due to non-Gaussian artifacts in the data. We suggest that
the autocorrelation test used here was instrumental towards achieving
our results and encourage the development and implementation of other
signal-based consistency tests which could be added to this analaysis
to improve upon these results (one such consistency test, known as the
bank veto~\cite{Hanna:2008}, is currently being tested within the {\tt
gstlal} pipeline). We also point out that the autocorrelation
consistency test is appealing from a computational point of view since
once a trigger has been produced by the pipeline, the needed SNR time
series is already available in memory and the calculation comes at
nearly no extra cost.

We have also taken a simplistic, but seemingly quite powerful,
approach to the matter of defining the coincidence of triggers between
detectors. For coincidence in time, we follow the standard interval
approach, requiring that triggers occur within 3ms of each other after
correcting for maximum light travel time between the detectors. For
mass and spin coincidence, however, we require that triggers in each
detector have identical parameters. This choice is possible in the
{\tt gstlal} framework since the same template bank is used for all
detectors and all times in the analysis. Previous LIGO and Virgo
searches for compact binary coalescence have used template banks whose
parameters depend on the local power spectral density of a detector,
resulting in template banks which are different in each detector and at different times. In
the latter implementation, the coincidence criterion must allow for
some small mismatch in the trigger parameters from different
detectors. Recent searches using two-parameter non-spinning template
banks have achieved this tolerance using estimates of the expected
uncertainty in parameter recovery to define a small error
region~\cite{ref:ethinca}, but the generalization of this technique to
higher dimensional parameter spaces is not straightforward and the
exact size of the error region typically requires careful tuning in
order to be effective. On the other hand, the exact parameter
coincidence feature of the {\tt gstlal} search pipeline generalizes
trivially to higher dimensional parameter spaces and requires no
tuning. The results here suggest the exact coincidence criterion is a
strong discriminator between background and signal, but we do not
systematically examine the relative merits of these two approaches.

Ultimately, the key to improving the sensitivity of a search pipeline
by the inclusion of more physical effects in the search templates is
the ability to manage the background trigger rates while exploiting
the elevation of the signal. The methods described here have proven
successful in mitigating the background elevation relative to the
signal to obtain a net gain in sensitivity. We have highlighted in
this section just two features of the {\tt gstlal} pipeline which are
manifestly different from other studies and lie at the core of the
background rejection techniques currently implemented in the
pipeline. Given that the {\tt gstlal} pipeline has not previously been
used for an analysis of this type, there are of course many other
differences between this work and previous studies, but isolating the
the particular features which made these results possible is a
difficult task.

\begin{figure*}

        \centering
        \subfloat[$\chi = 0$ Template Bank \label{fig:nonspin_mchirp_recovery}]{%
                \includegraphics[width=0.33\textwidth]{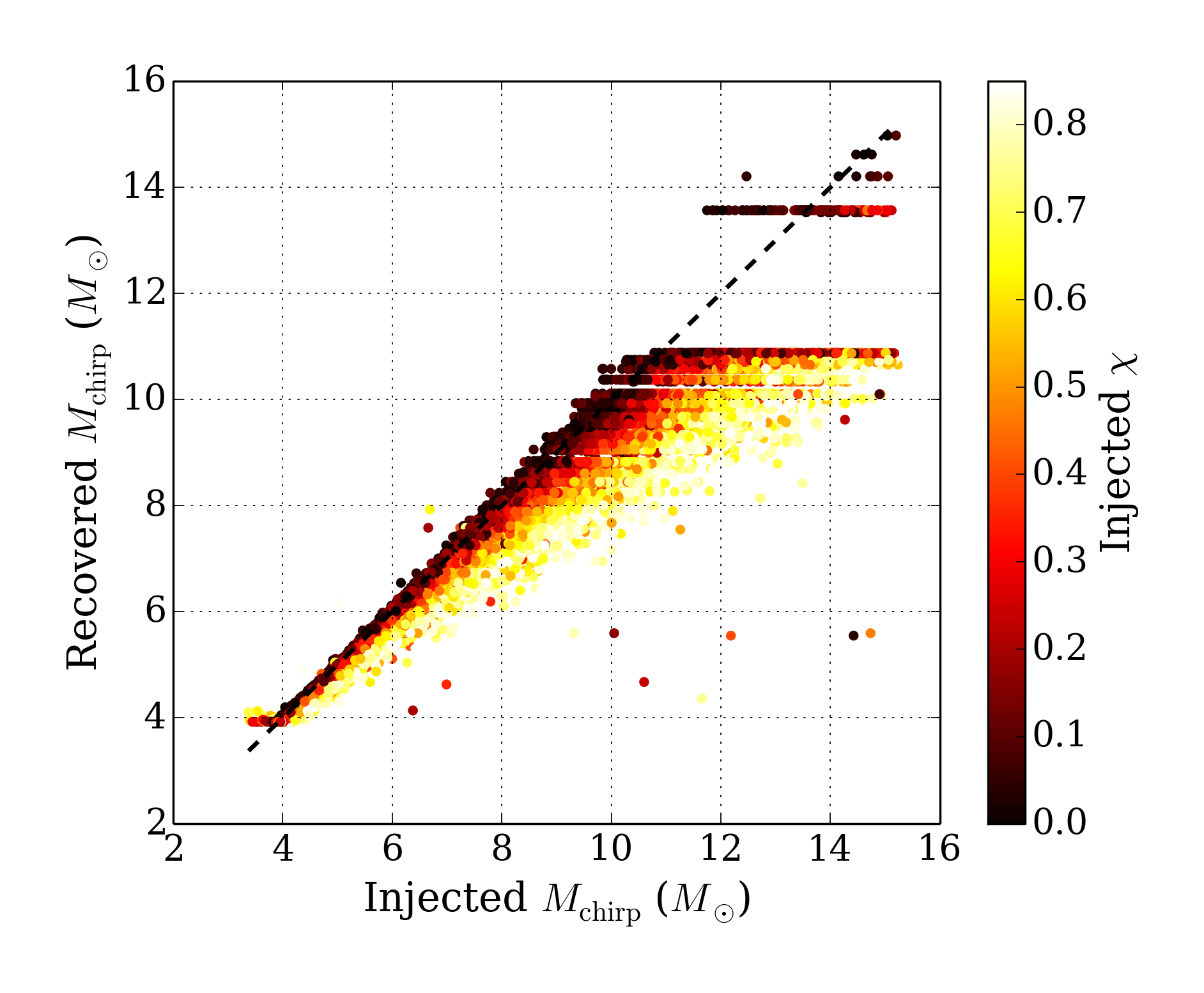}}
        \subfloat[$\chi \ge 0$ Template Bank \label{fig:positive_spin_mchirp_recovery}]{%
                \includegraphics[width=0.33\textwidth]{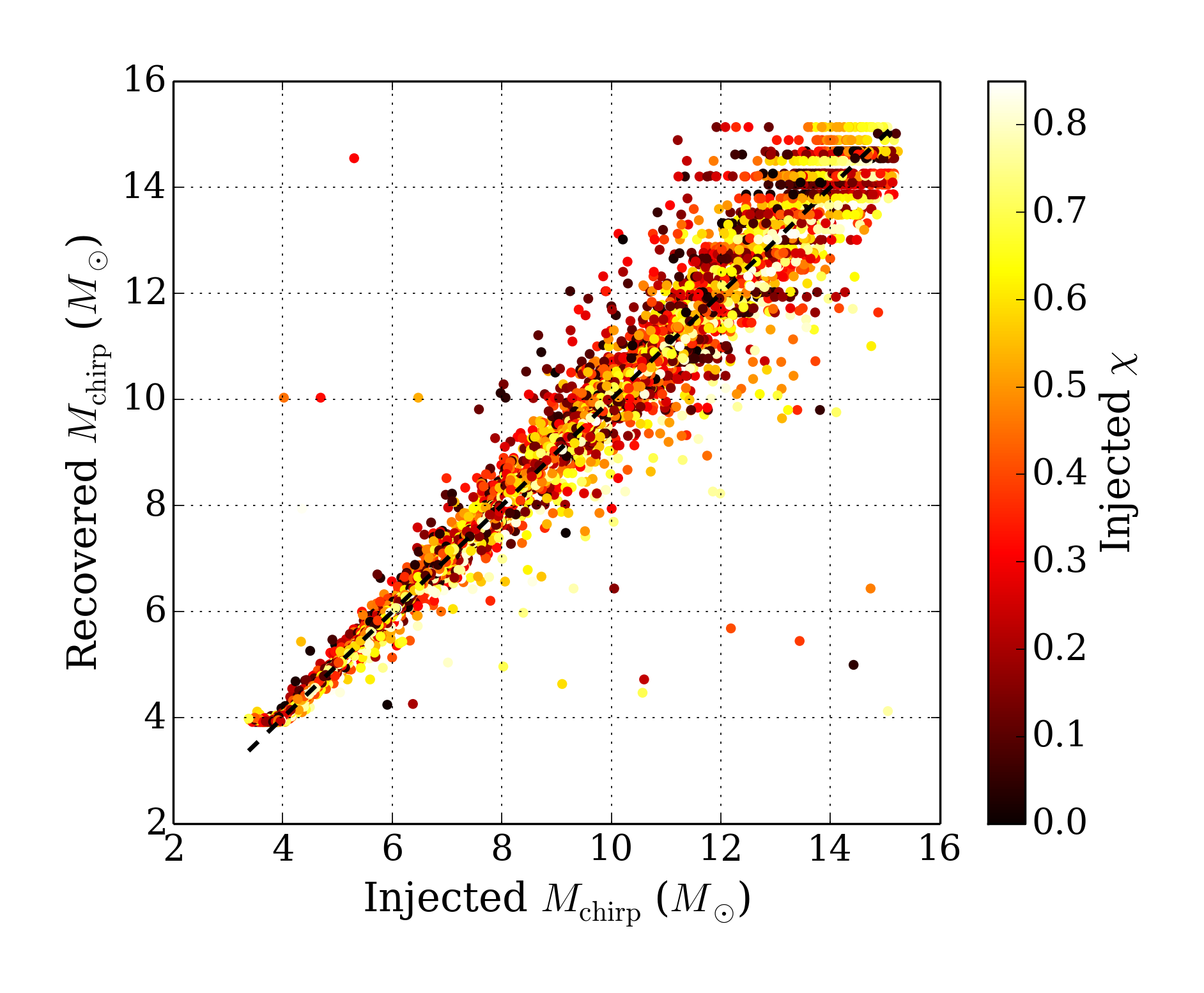}}
        \subfloat[$\chi \ge 0$ Template Bank \label{fig:positive_spin_spin_recovery}]{%
                \includegraphics[width=0.33\textwidth]{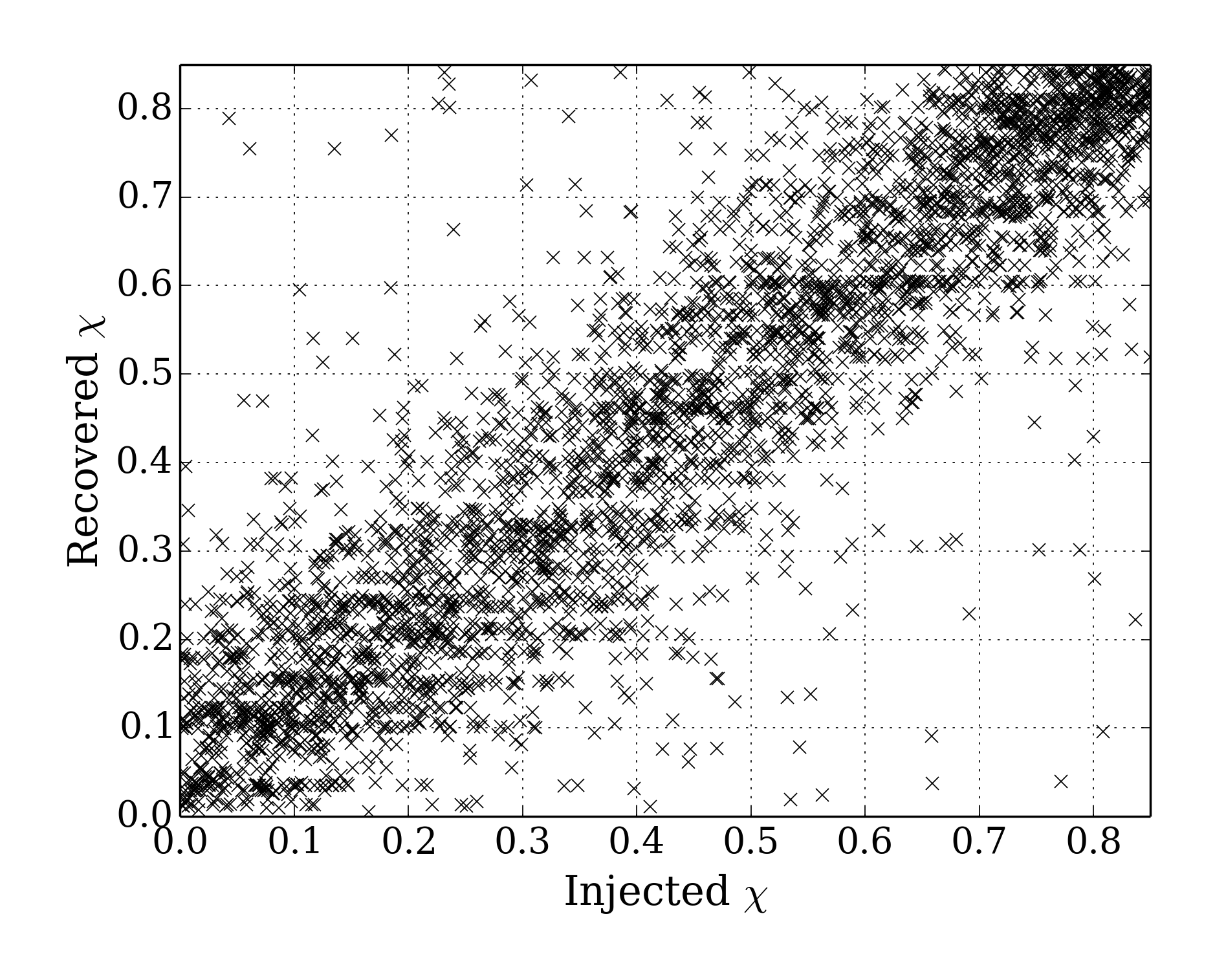}}

        \caption{\textit{Improvement in parameter recovery.}  In
          addition to improving the sensitive search volume, the
          inclusion of spin effects helps to curtail systematic biases
          in the recovery of template parameters. Here we demonstrate
          that the improvement in chirp mass recovery when spin is
          included in the templates. The left panel corresponds to the
          analysis with non-spinning templates while the middle and
          right panels corresponds to the analysis with aligned-spin
          templates.}

\label{fig:parameter_recovery_mchirp}
\end{figure*}

\section{Parameter Recovery}

There are robust, dedicated parameter estimations algorithms which can
extract the parameters of a signal, in gravitational wave data with
high accuracy, after the signal is found in a detection
algorithm~\cite{s6_pe}. These algorithms are computationally
intensive. Parameters of a signal can be inferred rapidly, albeit with
less accuracy, from a matched filter detection algorithm from the
identified template parameter.  In this paper, although we focus on
the detection performance, we also report two parameter extraction
properties of the search algorithm so as to understand any possible
biases in the search algorithm.
In Fig.~\ref{fig:nonspin_mchirp_recovery}, we show the
bias introduced in the recovery of the chirp mass parameter when
signals from positively aligned systems are searched for using
non-spinning templates. The bias indicates a degeneracy in the
parameter space. Positively aligned spins allow the binary to evolve 
to much higher frequencies (\textit{orbital
hang-up}~\cite{Campanelli:2006uy}) and this increases the length
of the waveform, as compared to its non-spinning counterpart. Thus,
high mass positively spinning systems may match best with lower mass
non-spinning systems. However, as shown in
Fig.~\ref{fig:positive_spin_mchirp_recovery}, when the proper spin
effects are included in the templates, the parameter bias is
substantially decreased. Furthermore, as seen in
Fig.~\ref{fig:positive_spin_spin_recovery}, we find that we can
recover the injected effective spin parameter with moderate accuracy, when we
include the spin in the search templates.

\section{Conclusion}
We have demonstrated the first implementation of a detection pipeline
for coalescing binary black holes which uses spinning templates to
achieve in real detector noise a greater sensitivity to spinning
signals. This analysis used the IMRPhenomB waveform model which
includes the full inspiral, merger and ringdown phases of the
coalescence and additionally models aligned-spin effects with a single
effective spin parameter $\chi$. Using this model, we constructed
two template banks of binary black holes with $M_\mathrm{total}\in
[10,35]M_\odot$ and $1\leq m_1/m_2 \leq 4$. In one template bank, we
restricted the templates to effective spins with $\chi=0$ and in the
other we included templates with $\chi \geq 0$. These banks were
constructed using the stochastic placement infrastructure {\tt SBank}
and implemented for filtering in the {\tt gstlal} search pipeline. We
measured and compared the sensitivity of this pipeline with each of
the constructed template banks to IMRPhenomB aligned-spin target
signals with $M_\mathrm{total} \in [15,25]M_\odot$ and $\chi \geq
0$. This sensitivity analysis was conducted with simulated signals
added to 25.9 days of real detector noise obtained from the fifth
observation run of the Hanford and Livingston detectors.

Our analysis showed an increase sensitive volume of up to 45\% for
target systems with $0.2 \leq \chi \leq 0.85$ when aligned-spin
effects are included in the templates. We have also reported that with
the use of aligned-spin templates in the pipeline, there is
added advantage that the effective spin parameter of the binary black
holes can be inferred and more accurate mass parameter estimation can
be achieved. This study is the first demonstration of a full detection
pipeline in which the use of spinning templates has been shown
increase the sensitivity to aligned-spin signals relative to the same
pipeline using non-spinning templates. We emphasize that this
improvement is demonstrated in real LIGO detector noise and therefore
includes the effects of dealing with of non-Gaussian noise artifacts.

The demonstration given here is encouraging and represents an
important step towards required to fully integrate spin effects in
LIGO and Virgo searches for compact binary coalescences. In this work,
we restricted our sensitivity measurements to binary black holes with
non-negative effective spins and total masses in $[15, 25]M_\odot$, a
very small, though astrophysically favored, portion of the much larger
parameter space accessible by LIGO and Virgo observations. Future work
will address some of these other regions of the mass and spin
parameter space and examine the noise-curve dependency of our
results. In particular, we plan to study the importance of spin for
searches of lower mass compact binary systems ($M< 12~M_\odot$). Such
systems may be well-modeled by inspiral-only waveforms and in this
regime there exist accurate waveform models for precessing binaries,
which may be particularly important for binary black holes. We expect
that when the sensitivities of the detectors are such that the lower
frequency can be profitably lowered from 40 Hz, as anticipated for the
advanced LIGO and Virgo detector era, the gains in sensitivity to
spinning signals achievable through the methods presented here will be
substantially greater.

\section{Acknowledgments}

The authors gratefully acknowledge the support of the United States
National Science Foundation for the construction and operation of the
LIGO Laboratory. This work was supported by NSF grants PHY-0855494 and
PHY-1207010.  PA's research was partially supported by a FastTrack
fellowship and a Ramanujan Fellowship from the Department of Science
and Technology, India and by the EADS Foundation through a chair
position on ``Mathematics of Complex Systems'' at ICTS-TIFR. The
authors would like to thank those individuals in the gstlal group who
wrote the programs used to perform the spinning analyses and Thomas
Dent for useful comments on the manuscript. SRPM would like to thank
Prayush Kumar for useful discussion. SP would like to thank Deborah
Hamm for useful discussion. This document has been assigned LIGO
laboratory document number P1200132.

\bibliography{paper}
\end{document}